# The reduced total isotope effect and its implications on the nature of superconductivity in MgB$_2$

D.G. Hinks[*], H. Claus[*, †], and J.D. Jorgensen[*]

[*]*Materials Science Division, Argonne National Laboratory, Argonne IL 60438*

[†]*Department of Physics, University of Illinois at Chicago, Chicago IL 60607*

The recent discovery of superconductivity at ~39 K in MgB$_2$, by Nagamatsu et al.[1] establishes this simple binary compound as having the highest bulk superconducting transition temperature, $T_c$, of any non-copper-oxide material. Much of the initial research has focused on whether MgB$_2$ is a conventional BCS, electron-phonon mediated superconductor, and, if it is, why $T_c$ is so high. Isotope effect measurements, in which the atom masses are manipulated to systematically change the phonon frequencies, are one of the fundamental experimental tests of electron-phonon mediated superconductivity. One would expect the total isotope effect coefficient, i.e. the sum of both the Mg and B coefficients, to be 1/2 for a high- $T_c$, phonon mediated, simple sp orbital superconductor like MgB$_2$. We find a value of 0.32(1), much reduced from the canonical BCS value of 0.5. This requires large values of $\lambda$ (electron-phonon coupling constant) and $\mu^*$ (repulsive electron-electron pseudopotential) to account for the high $T_c$ and, thus, constrains any theoretical model for superconductivity in MgB$_2$.

The search for superconductivity in transition metal diboride compounds is not new. In 1970, Cooper et al.[2] were able to obtain transition temperatures above 11 K in Zr$_{0.13}$Mo$_{0.87}$B$_2$. They also observed superconductivity in NbB$_2$. However, as this and other searches[3] showed, superconductivity in this class of compounds is not common and finding superconductivity near 39K in MgB$_2$ was certainly unexpected.

The presence of an isotope effect is a strong indicator of phonon mediation of the superconducting coupling. The large isotope effect measured by Bud'ko et al.[4] for B ($\alpha_B = 0.26(3)$) clearly shows that phonons associated with B vibrations play a significant role in the superconductivity of MgB$_2$. The isotope effect coefficient, $\alpha$, is defined as

$$\alpha = -\ d\ln T_c\ /\ d\ln M$$

for a one component system, where M is the atomic mass, and as

$$_i = -d\ln T_c /d\ln M_i$$

for each component, i, in a multicomponent system.  The $MgB_2$ system is unique in that both elements have isotopes with isotopic mass differences that could lead to a similar change in $T_c$ on substitution.  $MgB_2$ is one of the few compounds for which the total isotope coefficient, $_t$, can be accurately measured.

The crystal structure of $MgB_2$, composed of B graphite-like hexagonal nets separated by hexagonal close-packed Mg atom layers, is electronically simple, containing only bands formed from s and p orbitals.  Electronic band structure calculations by Kortus et al.[5] for $MgB_2$ show that the Mg donates substantially both of its 3s electrons to the B layer and the Fermi surface is derived mainly from B orbitals.  This leads to an ionic model of the material, $Mg^{+2}B_2^-$, and the superconductivity is, in their view, due to the "metallic" boron.  An and Pickett[6] agree with the ionic model in that Mg only acts as an electron donor, but they attribute the high $T_c$ to the two-dimensional nature of the boron Fermi surface.  In their model, charge is transferred from the $(sp^2$ orbitals) to the $(p_z$ orbitals) to create hole carriers in the strongly bonding bands.  They conclude that the strong coupling of the holes to the in-plane B phonon lattice vibrations ($E_{2g}$ modes) is responsible for the high $T_c$.

For our study, the $MgB_2$ samples were made from Eagle-Picher isotopic B ($^{11}$B and $^{10}$B enriched to 99.52% and 99.49%, respectively) and Oak Ridge Mg isotopes ($^{24}$Mg and $^{26}$Mg enriched to 99.92% and 99.44%, respectively).  The sample size was small, about 0.1g, because of the high cost of the isotopic Mg.  The Mg (as small pieces) and B (<100 mesh powder) were weighed into small machined BN ( Advanced Ceramics Corp. grade HBC) crucibles with closely fitting lids.  The samples were fired at 850$^O$C for 1.5 hours under 50 bar of ultra-high-purity Ar.  There was some discoloration of the BN crucibles but no major reaction.  Six different samples were made using combinations of the 3 Mg isotopes (24, 26 and natural) and $^{11}$B and $^{10}$B.

The magnetizations of the samples were measured in a noncommercial, low-field superconducting quantum interference device magnetometer, which was previously used extensively in the characterization of superconducting samples[7].  Approximately 5 to 10 mg of powder was contained in a gelatin capsule for each measurement.  This magnetometer employs a Cu solenoid for fields between 0 and 100 G.  With the help of a double μ-metal shield, the residual field was reduced to about 2 mG.  The magnetization data were taken after cooling in zero applied field to about 20 K.  A magnetic field of 1 G was then applied and the magnetization was recorded on warming.

Figure 1a shows the magnetization data through the superconducting transitions for the six different isotopic samples. It is immediately obvious that the B isotope effect is large and that Mg shows only a small effect. The B isotope effect was determined using both the extrapolated onset $T_c$ and a 10% $T_c$ criterion (i.e. the temperature at which the signal is 10% of the full diamagnetic signal). Table 1 shows both the onset and 10% criteria $T_c$ for the six samples and the calculated values of $\alpha_B$. Both sets of data agree within experimental error giving a value of 0.30(1) for $\alpha_B$, within the experimental error of the value reported by Bud'ko et al.[4] of 0.26(3). The value of $\alpha_B$ found for $MgB_2$ is quite close to the B isotope effect found in the borocarbides: $YNi_2B_2C$, 0.25(4)[8], $YPd_5B_3C_{0.3}$, 0.32(4)[8], $LuNi_2B_2C$, 0.11(5)[9]. Cheon et al.[9] estimated that the total isotope effect for these material should be near 0.5, but the lack of suitable isotopes (Y) or heavy masses (Lu) make experimental verification difficult.

The Mg isotope effect is very small, but non zero, as illustrated in Figures 1b and c with an expanded temperature scale. The [10]B and [11]B isotopic samples clearly show a Mg isotope effect of the expected sign; however, the natural abundance Mg samples, labeled [n]Mg, show the highest $T_c$'s. This may be due to some small impurity content in the Mg isotopes, suppressing the $T_c$'s of these samples relative to the natural, high purity Mg sample. Since it is impossible to quantify the impurity effects in this measurement, all three Mg isotope masses were used for each B isotope to calculate $\alpha_{Mg}$. Figure 2 shows the least squares fit of the transition temperatures vs. Mg masses for both B isotopes using both the onset and 10% $T_c$ criteria. Averaging both the onset and 10% $T_c$'s, and the values for both B isotopes, $\alpha_{Mg}$ is found to be 0.02(1).

$T_c$ for a superconductor can be written in a closed form as a product of a weighted average over the phonon frequency spectra $<\omega>$ times a function of the coupling constant ($\lambda$) and the coulomb repulsion pseudopotential ($\mu^*$), and is generally approximated by the Allen-Dynes[10] modification of the McMillan equation[11]

$$T_c = \langle \omega \rangle e^{-f(\lambda,\mu^*)} \quad \frac{\omega_{ln}}{1.2} e^{-\frac{1.04(1+\lambda)}{\lambda-\mu^*(1+0.62\lambda)}}$$

where $\omega_{ln}$ is a characteristic phonon energy determined by a weighting as defined in reference 9. For a simple one component, phonon-mediated BCS superconductor with $\mu^* = 0$, this form of the equation for $T_c$ will give

$$T_c \propto \ln \lambda \, (1/M)^{1/2}$$

since $\lambda$ is, in general, thought to be mass independent. The isotope coefficient under these assumptions is 1/2. However, it may vary greatly if $\mu^* \neq 0$ and $\lambda$ is small (as is, for example, the case for some low $T_c$ materials). Garland[12] has shown that for transition metals, Coulomb effects reduce $\beta$ due to the presence of d orbitals whereas sp metals generally show a nearly full isotope effect. Using the approximate form of the McMillan equation shown above, $\beta$ can be written[11,13] in a closed form as a function of $\lambda$ and $\mu^*$,

$$\beta = \frac{1}{2}\left[ 1 - \frac{1.04(1+\lambda)(1+0.62\lambda)\mu^{*2}}{\left\{\lambda - \mu^*(1+0,62\lambda)\right\}^2} \right]$$

which reduces to 1/2 for $\mu^* = 0$.

In the case of a multicomponent system, $\beta_t$, the total isotope effect coefficient, which is the sum of all individual isotopic effect coefficients, would tend to 1/2 at low values of $\mu^*$ and large values of $\lambda$.[13] In this case, the functional form of the averaging process over the phonon density of states to obtain $<\omega>$ will determine the individual isotope effect coefficients, $\beta_i$. Recently, Osborn et al.[14] have calculated $\beta_B$ and $\beta_{Mg}$ using a Born-von-Karman model verified by comparison with the measured phonon density-of-states. Depending on their weighting of the phonon density of states, they obtain 0.42 and 0.08 or 0.28 and 0.22 for $\beta_B$ and $\beta_{Mg}$, respectively. As expected, the individual isotope effect coefficients varied widely depending on the exact functional form of the prefactor, $<\omega>$, for the $T_c$ equation, however, the total isotope coefficient remained at 1/2.

MgB$_2$ appears to be a conventional, BCS phonon-mediated sp superconductor with an intermediate-to-strong electron-phonon coupling strength. Specific heat measurements[15] and $^{11}$B nuclear spin lattice relaxation measurements[16] indicate a moderately strong coupling superconductor. The superconducting gaps determined from tunneling measurements range from 2meV to 7 meV[17-20], spanning the range of weak to moderate coupling strength (5.9meV is the weak coupling limit gap value). There is some evidence that the low gaps result from a weakened surface layer[19], thus, if the largest gaps are intrinsic to the material, tunneling measurements would place MgB$_2$ in the moderately strong limit. Estimates of $\lambda$ range from 0.65 to 1.2 with values of $\mu^*$ ranging from near 0 to 0.2.[5,14,21]

The measured reduced total isotope coefficient of 0.32 could be the result of a large coulomb pseudopotential, Our measured value can place some limits on the values of   and μ*.  Figure 3 shows the allowable range of μ* for different  's calculated with the equation shown above for   .  μ* must, of necessity, be larger then the canonical value of 0.1 generally assumed for this variable in order to account for the reduced total isotope coefficient.  The results shown in figure 3 place demanding constraints on theoretical attempts to explain the high $T_c$.  For example, the recent electronic structure calculations for $MgB_2$ by Kong et al.[19] resulted in a value of      0.65, requiring a value of μ* of about 0.02 to account for the observed $T_c$ of 39K.  These values of    and μ* would lead to a total isotope coefficient near 0.5, casting some doubt on their model of the $MgB_2$ system.  $T_c$ can also be calculated if one knows the value of   $_{ln}$.  Using the value of 57.9 meV determined by Osborn et al.[14], the calculated $T_c$ 's as a function of     are also show in Figure 3.  The values found for    and μ* for a $T_c$ of 39K, about 1.4 and 0.25 respectively, are larger than most estimates so far determined for these parameters.

Our measurement of the Mg and B isotope effects shows that there is substantial coupling of B atom vibrations to the electronic structure.  The very small value of 0.02(1) for   $_{Mg}$ show that vibrational frequencies of Mg atoms have almost no effect on $T_c$, however, one cannot rule out the possibility that the Mg vibration modes still contribute to $T_c$.  The presence of an isotope effect clearly indicates a phonon contribution to $T_c$, but the absence or measurement of a small value is ambiguous.  It is only the total isotope effect that has any quantative meaning within the context of simple BCS superconductivity.  We have argued that our observed reduced total isopope coefficient is the result of a large coulomb repulsion resulting in a large value of μ*.  Certainly, many other causes for a reduced total isotope effect can occur, e.g., anharmonic lattice vibration could effect the mass dependence of < > or complex dimensional effects on the Fermi surface could effect f( ,μ*).  The model of An and Pickett[6], for example, might lead to a reduced total isotope effect if the two-dimensional nature of the hole band is found to have an effect on    or μ*.  This difference between the measured total isotope effect and 1/2 may appear insignificant, but the cause of this discrepancy may be in some way related to the high $T_c$ observed in this material.  Although it is possible to calculate $T_c$ for $MgB_2$ based on reasonable choices of parameters using the McMillan equation, correctly modeling   $_t$ (and   $_{Mg}$ and   $_B$) may place constraints on these calculations that will provide important insight into how such a high $T_c$ is achieved in this simple sp metal.

**Acknowledgements**
This work was supported by the U.S. Department of Energy, Office of Science.

**Correspondence and requests for materials should be addressed to D.G.H. (e-mail: hinks@anl.gov).**

| B Mass | Mg Mass | $T_c$ (K) Onset | $T_c$ (K) 10% | B Onset | B 10% |
|--------|---------|-----------------|---------------|---------|-------|
| 10.0051 | 25.001 | 40.21(2) | 40.16(2) | 0.31(1) | 0.31(1) |
| 10.9952 |  | 39.06(2) | 39.00(2) |  |  |
| 10.0051 | 24.305 | 40.23(2) | 40.21(2) | 0.29(1) | 0.30(1) |
| 10.9952 |  | 39.16(2) | 39.09(2) |  |  |
| 10.0051 | 24.001 | 40.25(2) | 40.20(2) | 0.30(1) | 0.31(1) |
| 10.9952 |  | 39.12(2) | 39.06(2) |  |  |
|  |  |  | average | 0.30(1) | 0.31(1) |

Table 1. Onset and 10% $T_c$ values for the six different isotopic $MgB_2$ samples. B is calculated for each of the Mg isotopes for the two $T_c$ criteria.

Figure Captions

Figure 1. The superconducting transitions for the isotopically substituted $MgB_2$ samples. Figure 1a shows the relative magnetization for all samples. $^nMg$ indicates samples with natural Mg. Figures 1b and 1c illustrate the small Mg isotope effect on an expanded temperature scale for the $^{11}B$ and $^{10}B$ samples, respectively.

Figure 2. The superconducting transition temperatures for samples with different Mg isotopic masses for the $^{10}B$ and $^{11}B$ substituted samples. The onset $T_c$ (solid lines) and 10% $T_c$ (dashed lines) are least squares fits to the data.

Figure 3. The allowable values of $\mu^*$ (shown in the top crosshatched area) as a function of based on the McMillan equation for $T_c$ if $_t$ ranges from 0.30 to 0.34. The $T_c$ of $MgB_2$ (shown in the lower crosshatched area) as a function of for the allowable values of $\mu^*$ using $_{ln}$ = 57.9 meV.

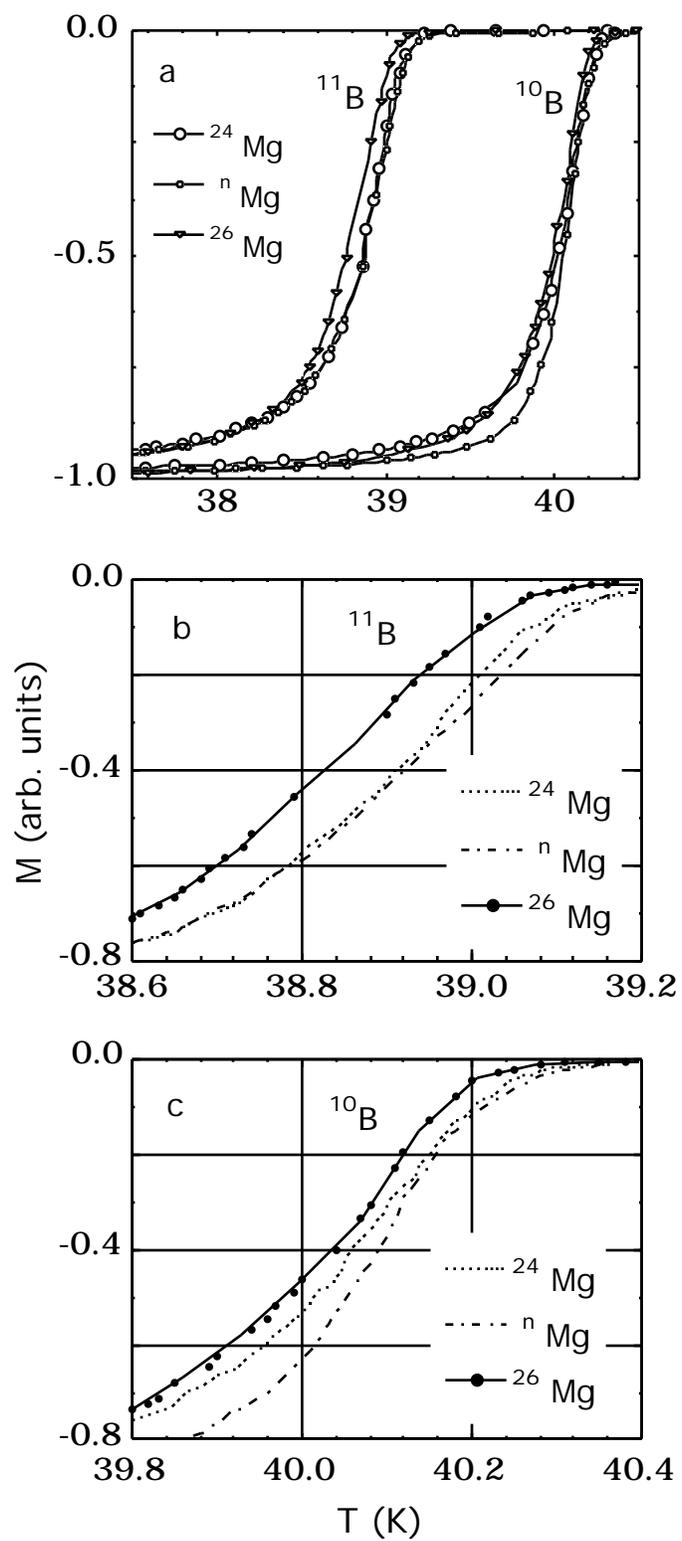

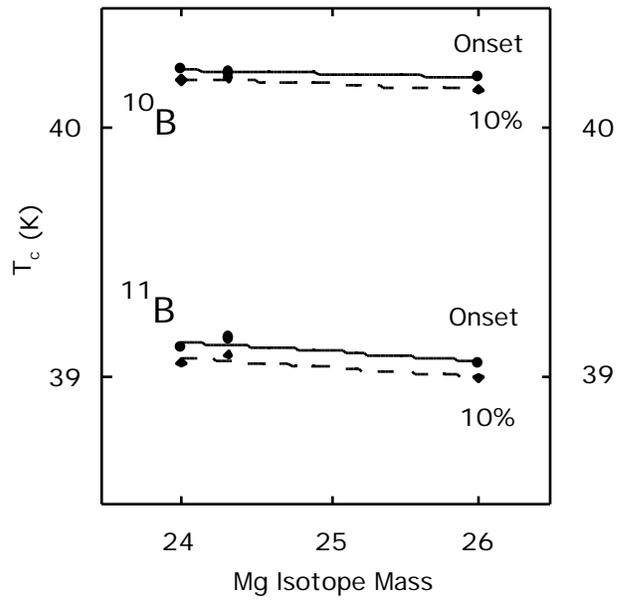

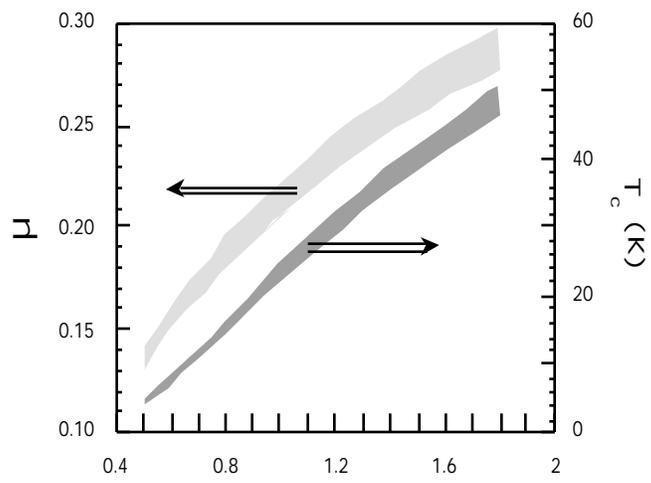